\documentclass[letterpaper,twoside,twocolumn]{IEEEtran}
\usepackage[T1]{fontenc}
\usepackage[latin9]{inputenc}
\usepackage{amsmath}
\usepackage{graphicx}
\usepackage{amssymb}
\usepackage{esint}

\usepackage[english]{babel}

\begin{document}
\selectlanguage{english}

\title{Analysis of Uncoordinated Opportunistic Two-Hop Wireless Ad Hoc Systems}

\author{Radha Krishna Ganti and Martin Haenggi\\
Department of Electrical Engineering\\
University of Notre Dame\\
Notre Dame, IN 46556, USA\\
\{rganti,mhaenggi\}@nd.edu}
\maketitle
\begin{abstract}
We consider a time-slotted two-hop wireless system in which the sources
transmit to the relays in the even time slots (first hop) and the
relays forward the packets to the destinations in the odd time slots
(second hop). Each source may connect to multiple relays in the first
hop. In the presence of interference and without tight coordination
of the relays, it is not clear which relays should transmit the packet.
We propose four decentralized methods of relay selection, some based
on location information and others based on the received signal strength
(RSS). We provide a complete analytical characterization of these
methods using tools from stochastic geometry. We use simulation results
to compare these methods in terms of end-to-end success probability.
\end{abstract}

\section{Introduction}

We consider a two-hop wireless ad hoc network in which the sources
are distributed randomly on the plane and each source has a destination
at a distance $R$ in a random direction. In addition there exists
a set of relays (different from the destinations) which assist the
sources. In the first hop each source transmits and will be decoded
by any node listening (relays and destinations) with signal-to-interference
ratio (SIR) greater than a fixed threshold $T$. In the second hop,
some of the relays which were able to decode information in the first
hop, transmit. So in the first hop each packet may be received by
many relays, hence multiple copies of the same packet may exist at
different relays. In networks with low or no mobility multiple copies
may be avoided by setting the routing tables apriori and relays rejecting
packets if the source-destination pair is not in its routing table.
Without routing tables, the relay nodes having the same packet may
have to coordinate with each other and then choose one among them
to transmit the packet. In a mobile wireless network such a coordination
incurs significant overhead and also restrict the number of relays
per source-destination pair to one. More importantly such a restriction
\emph{may} in effect reduce the probability of success. It is not
clear how to choose a subset of these intermediate relays in a distributed
fashion so as to reduce the interference and increase the probability
of packet delivery. In this paper, we analyze the success probability
of such a two-hop scheme taking the interference and the spatial statistics
of the transmitting nodes into account. 

\emph{Related Work:} In \cite{cui2007orw}, relay selection is based
on the SINR. All the relay-destination channel states are assumed
to be known at the destination, and the destination chooses the relay.
In \cite{lo2007ors} the relay selection is based on the channel state
information that is fed back to the source. In \cite{bletsas2005sdm}
the relays estimate the channel using channel reciprocity theorem
and use timers to select the best relay. In \cite{zorzi2003} a relay
selection method (GeRaF) based on the distance from the destination
is considered. TDMA based contention is used to resolve the relays
transmitting to the same destination. In all these method some form
of channel contention and feedback is used to select the relays. In
the methods we analyze relays are selected in a completely distributed
fashion without any channel feedback. This makes these relay selection
schemes suitable in scenarios with moderate to high mobility.

The main contribution of this paper is the complete analytical description
of a two-hop wireless network with interference when the nodes and
relays are distributed as a Poisson point process on the plane. The
techniques developed in this paper can be used to analyze other position
based relay selection methods.

\section{System Model}

We assume that the sources form a Poisson point process (PPP) $\phi_{s}$
of intensity $\lambda_{s}$ on the plane. The relays are also assumed
to form a PPP $\phi_{r}$ of intensity $\lambda_{r}$ on the plane.
Each source $x\in\phi_{s}$ has a destination denoted by $r(x)$ (not
a part of $\phi_{s}$ or $\phi_{r}$) at a distance $R$ in some random
direction. We assume that the fading between any two nodes is Rayleigh
distributed so that the powers are exponential with unit mean. A transmitter
located at $x$ can communicate with a receiver at $y$ if $\text{SIR}(x,y\mid\phi)>T$.
The SIR is defined as \[
\text{SIR}(x,y\mid\phi)=\frac{h_{xy}g(x-y)}{\sum_{z\in\phi\setminus\left\{ x\right\} }h_{zy}g(z-y)}\]
where $\phi$ is the transmitting set, $g(x)$ is the path loss function,
and $h_{xz}$ is the power fading coefficient between nodes $x$ and
$z$. We assume $T>1$, i.e., a narrow band system which implies at
most one transmitter can connect to any receiver. The path loss function
$g(x)$ is assumed to depend only on $\|x\|$, to monotonically decrease
with $\|x\|$, and $\lim_{a\rightarrow\infty}g(a)a^{2}=0$ to guarantee
finite mean interference We restrict the number of hops between any
source-destination pair to be two. So a source can reach its destination
either in a single hop or can use the relays to reach the destination.
We can assume that the sources transmit in the even time slots and
a subset of the relays in the odd time slots. \emph{}\\
\emph{Notation: }
\begin{itemize}
\item We define \[
\mathbf{1}(x\rightarrow y\mid\phi)=\mathbf{1}(\text{SIR}(x,y\mid\phi)>T).\]
 $\mathbf{1}(x\rightarrow y\mid\phi)$ represents a random variable
that is equal to one if a transmitter at $x$ is able to connect to
a receiver $y$ when the transmitting set is $\phi$.
\item We define for $x\in\phi_{s}$ \[
\hat{\phi}_{r}(x)=\left\{ y\in\phi_{r}:\ \mathbf{1}(x\rightarrow y\mid\phi_{s})\right\} .\]
 $\hat{\phi}_{r}(x)$ denotes the cluster of relays to which the source
$x$ is able to connect in the first hop. 
\end{itemize}
\emph{Metric:} We analyze the success probability for the direct connection
between the source-destination and the two-hop connection between
them separately. Let $P_{1}$ denote the probability that a source
can connect to its destination directly in the first hop. More precisely
we define \begin{equation}
P_{1}=\lim_{a\rightarrow\infty}\frac{1}{\lambda_{t}\pi a^{2}}\mathbb{E}\sum_{x\in\phi_{s}\cap B(o,a)}\mathbf{1}(x\rightarrow r(x)\mid\phi_{s}).\label{eq:p1}\end{equation}
 The relays which were able to connect to some source in the first
hop are the potential transmitters in the second hop. In the relay
selection methods studied in the next section, a subset of these potential
transmitters $N_{x}\subseteq\hat{\phi}_{r}(x)$ are selected for each
$x\in\phi_{s}$ to transmit in the next hop. Let the probability that
a relay can connect to its intended destination (determined by the
source to which it connects in the first hop) in the second hop be
$P_{2}$, defined as \begin{eqnarray}
 &  & P_{2}=\lim_{a\rightarrow\infty}\frac{1}{\lambda_{t}\pi a^{2}}\mathbb{E}\sum_{x\in\phi_{s}\cap B(o,a)}1-f(x),\label{eq:p2}\end{eqnarray}
where $f(x)=\prod_{y\in N_{x}}\left(1-\mathbf{1}(y\rightarrow r(x)\mid\psi)\right)$
and $\psi=\cup_{x\in\phi_{s}}N_{x}$. In the above equation $1-f(x)$
is equal to one if and only if at least one node belonging to $\hat{\phi}_{r}(x)$
is able to connect to the destination $r(x)$. Here we are assuming
no cooperative communication between nodes which have the same information,
so relays belonging to the same cluster $\hat{\phi}_{r}(x)$ also
interfere with each other in the second hop. Then the success probability
is given by 

\[
P_{s}=1-(1-P_{1})(1-P_{2}).\]
 In the above equation we assumed that the success probability of
the direct connection is independent of the two hop success probability.
We also used the spatial ergodic property of PPP in defining $P_{1}$
and $P_{2}$.

\section{Location-Unaware Relay Selection }

In all the four methods described below, a relay has to make a decision
whether to transmit in the second hop if it is able to connect to
some source in the first hop. More precisely:
\begin{enumerate}
\item The destination $r(x)$ can directly decode a packet from $x\in\phi_{s}$
in the first hop if the $\text{SIR}(x,r(x)\mid\phi_{s})>T$.
\item A subset of the relays $N_{x}\subseteq\hat{\phi}_{r}(x)$ are chosen
in a distributed manner to transmit in the second hop.
\item In the second hop, the destination $r(x)$ can decode the packet from
a relay $y\in\phi_{r}$ if the $\text{SIR}(r(x),y\mid\psi)>T$.
\end{enumerate}
The success probabilities of all these methods is analyzed in the
Appendix.

\subsection{Method $1$: All relays transmit}

This is the most basic scheme where all relays which receive in the
first time slot transmit in the second hop: \[
N_{x}=\hat{\phi}_{r}(x).\]
As we shall see, this method has the worst performance because of
the high interference present in the second hop (specially when the
relay density is high).

\subsection{Method $2$: RSS-based selection}

When a relay node is able to connect to a source node, the relay node
has information about the RSS and could use that information to make
a decision to transmit in the next hop. In Method $1$, we do not
utilize any information regarding the SIR received at the relay. In
this method we utilize the RSS information to make the decision. We
have $\text{RSS}=S+I$ where $S$ is the strength of the desired signal
and $I$ is the interference observed. Since the relay was able to
decode the source, we have $\frac{S}{I}>T$ and thus \[
I\leq\frac{\text{RSS}}{1+T}\leq\frac{S}{T}.\]
Hence a small value of $\frac{\text{RSS}}{1+T}$ implies low interference
(and hence may see few interferer's in the second hop). This might
also mean smaller $S$ which implies that the relay is far from the
source. A large value of $\frac{\text{RSS}}{1+T}$ implies a large
$S$ and hence would indicate a relay close to the source. This may
potentially also imply more interference at the relay. So in the second
hop we give higher priority to nodes that observe smaller value of
$\frac{\text{RSS}}{1+T}$. A relay $y\in\hat{\phi}_{r}(x)$ transmits
the packet in the second time slot with probability \begin{equation}
\exp\left(-\delta\frac{\text{RSS}(y)}{1+T}\right),\label{eq:rss-selection}\end{equation}
where $\text{RSS}(y)$ is the RSS that the relay $y$ observes and
$\delta$ represents a parameter to be chosen so that $\frac{\delta\text{RSS}(y)}{1+T}$
is not too large. We have chosen an exponential penalty just for convenience
and the effect of the penalty function should be investigated further.
So we have \begin{equation}
N_{x}=\left\{ y\in\hat{\phi}_{r}(x):\ U_{y}<\exp\left(-\delta\frac{\text{RSS}(y)}{1+T}\right)\right\} \label{eq:method2}\end{equation}
where $U_{y}$ is a set of i.i.d uniform random variables in $[0,1]$.

We observe that Methods $1$ and $2$ are completely decentralized
and require no information about the location of any node. Hence these
methods of relay selection work perfectly even with high mobility
and incur zero overhead.

\section{Location-aware selection}

In the location-aware based methods we assume that each node has knowledge
about its own location and each source knows the spatial location
of its destination. Also each packet has information about the location
of the source from which it originated and the location of its destination
in its header.

\subsection{Method $3$: Sectorized relay selection}

\label{sub:sectorized-relay}

After a relay receives a packet in the first hop, it calculates the
angle between the relay-source and the source destination and makes
a decision based on this information. More precisely a relay $y\in\hat{\phi}_{r}(x)$
transmits the packet in the second time slot if $\angle yxr(x)<\theta$.
So for this method, we have \[
N_{x}=\left\{ y\in\hat{\phi}_{r}(x):\ \angle yxr(x)<\theta\right\} .\]
 In this method we are reducing interference by choosing relays in
a sector. If the angle $\theta$ is properly chosen the sector may
contain only one relay, i.e., $|N_{x}|=1$ which would eliminate the
intra-cluster interference.

\subsection{Method $4$ : Distance-based selection}

In this method, we select the relays in $\hat{\phi}_{r}(x)$ depending
on their distance from the destination $r(x)$. A relay $y\in\hat{\phi}_{r}(x)$
transmits the packet in the second time slot with probability \begin{equation}
\exp\left(-\epsilon\frac{2\|y-r(x)\|}{R}\right)\label{eq:distance-based}\end{equation}
 We have normalized the distance by $R/2$ so that nodes closer to
the source are not over penalized. $\epsilon$ is a normalizing parameter
which we will choose later. So we have \[
N_{x}=\left\{ y\in\hat{\phi}_{r}(x):\ U_{y}<\exp\left(-\epsilon2\|y-r(x)\|R^{-1}\right)\right\} \]
where $U_{y}$ is a set of i.i.d uniform random variables in $[0,1]$.
In this method, we observe that relays close to the destination have
a higher chance of transmitting than those closer to the source. One
could replace $\|y-r(x)\|$ in \eqref{eq:distance-based} with $\|y-x\|$
and $\exp$ with $1-\exp$. Then the source need not know its destination
location but one would loose the directionality of relay selection.

In the location-aware methods proposed in this section, each source
needs to know where its destination is located and maintaining this
information would require a significant overhead in a mobile network.

\section{Simulation Results}

In this section we compare the four methods described in the above
section by simulations. For the purpose of simulation we consider
a square $[-30,30]^{2}$ in which the nodes are located and use Monte-Carlo
method to evaluate the results. We use $T=3$ for all the simulations.

\begin{figure}
\begin{centering}
\includegraphics[width=3.1in]{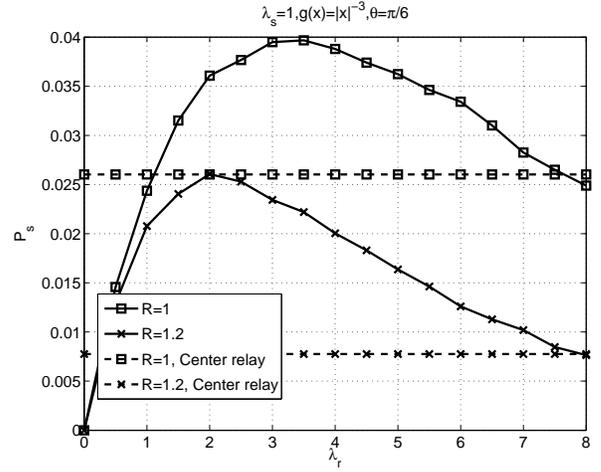}
\par\end{centering}

\caption{Comparison of Method $3$ versus the center-scheme. The success probability
versus the relay intensity $\lambda_{r}$ is plotted for various values
of the source destination distance $R$. }

\label{Flo:method-center}
\end{figure}

In Figure \ref{Flo:method-center}, we compare Method $3$ with a
scheme in which the relays are apriori chosen. In this scheme each
source-destination pair has one relay assisting in communication.
The relay is centered halfway between the source and the destination.
Intuitively such a scheme is the best single-relay scheme in terms
of end-to-end success probability. We observe that using uncoordinated
relays yields a better performance than selecting a relay apriori.
This is because of the selection diversity that occurs due to fading
and the node locations. Also observe that there is an optimal value
of relay intensity that achieves the maximum value of $P_{s}$. In
practice this can be achieved by starting with large number of relays
and using ALOHA-like thinning in the second hop. We observe that the
intensity of relays required to outperform the center-method decreases
with increasing source destination distance $R$. %
\begin{figure}
\begin{centering}
\includegraphics[width=3.1in]{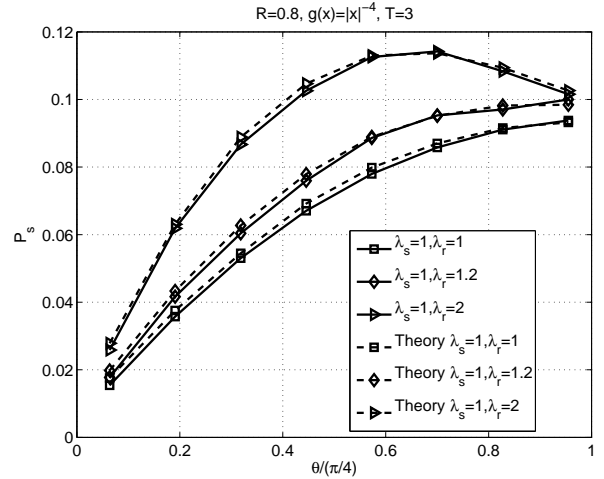}
\par\end{centering}

\caption{Comparison of the simulation and theoretical results for Method $3$.}

\label{Flo:theory-comp}
\end{figure}
From Figure \ref{Flo:theory-comp}, we observe that the probability
of success as computed from the theory matches  the simulations closely.
\begin{figure}
\begin{centering}
\includegraphics[width=3.1in]{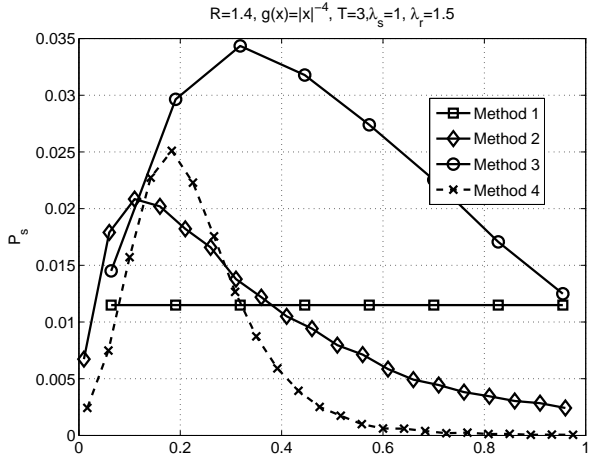}
\par\end{centering}

\caption{Probability of success for the four methods described in the previous
section. The $x$-axis represents $\delta$ for Method $2$, $4\theta/\pi$
 for Method $3$ and $\epsilon/6$ for Method $4$. }

\label{Flo:angle-method}
\end{figure}
In Figure \ref{Flo:angle-method}, the success probability is plotted
for the four methods described in previous sections. We first observe
that all the methods have an optimal value of the parameters $\delta$,
$\theta$ and $\epsilon$ that achieve the maximum success probability.
We also observe that Method $1$ has the worst performance. This is
because of the high interference caused by all the relays (that have
received a packet in the first hop) transmitting in the second hop.
We observe that Method $3$ i.e., the sectorized selection method
performs the best. Also observe that the RSS based selection has twice
the $P_{s}$ as compared to Method $1$ in which every relay transmits.
We also observe that the both the location-aware schemes outperform
Method $1$ and Method $2$ for the particular values of $T=3,\lambda_{s}=1,\lambda_{r}=1.5$.

\section{Conclusion}

In this paper, we have have analyzed the success probability in a
two hop system taking interference and spatial distribution of the
nodes into account. We have provided an analytical solution for all
the four methods using some approximations. This method of analysis
can be easily extended to any position based relay selection. We have
shown that uncoordinated selection of relays increases the success
probability as compared to selecting a relay for each source-destination
pair apriori.

\section*{Appendix}

\label{sec:proof}

The probability that a destination located at $z$ can decode the
packet transmitted by a source $\xi$ when the interference is caused
by $\phi_{s}$ is \begin{eqnarray}
p_{\xi}(z) & = & \mathbb{P}\left(\xi\rightarrow z\mid\phi_{s}\right)\label{eq:33}\\
 & = & \mathbb{P}\left(\text{SIR}(\xi,z\mid\phi_{s})>T\right)\\
 & = & \mathbb{P}(h_{\xi z}>\frac{T}{g(\xi-z)}\sum_{x\in\phi_{s}}h_{xz}g(x-z))\nonumber \end{eqnarray}
 \begin{eqnarray*}
 &  & \stackrel{(a)}{=}\mathbb{E}\prod_{x\in\phi_{s}}\exp\left(-\frac{T}{g(\xi-z)}h_{xz}g(x-z)\right)\\
 &  & \stackrel{(b)}{=}\exp(-\lambda_{s}\int_{\mathbb{R}^{2}}\beta(\xi-z,y)dy)\end{eqnarray*}
where \[
\beta(x,y)=\frac{1}{1+\frac{g(x)}{T}g(y)^{-1}}\]
$(a)$ follows from the exponential distribution of $h_{\xi z}$,
and $(b)$ follows the probability generating functional (PGFL) of
the PPP \cite{bacelli-aloha,stoyan}. Also observe that $p_{\xi}(z)$
depends only on $\|\xi-z\|$. 

\emph{Direct transmission}: So from \eqref{eq:p1} we have \begin{eqnarray*}
P_{1} & = & \lim_{a\rightarrow\infty}\frac{1}{\lambda_{t}\pi a^{2}}\mathbb{E}\sum_{x\in\phi_{s}\cap B(o,a)}\mathbf{1}(x\rightarrow r(x)\mid\phi_{s})\\
 & \stackrel{(a)}{=} & \lim_{a\rightarrow\infty}\frac{1}{\pi a^{2}}\int_{B(o,a)}p_{x}(r(x))dx\ \stackrel{(b)}{=}\ P_{o}(R),\end{eqnarray*}
where $(a)$ follows from the Campbell-Mecke theorem \cite{stoyan}
and $(b)$ follows from he fact that $p_{\xi}(z)$ depends only on
$\|\xi-z\|$. 

\emph{First hop}: A point process is completely characterized by its
PGFL and so we will evaluate the PGFL of the relays which can connect
to source $\xi\in\phi_{s}$, i.e., the cluster $\hat{\phi}_{r}(\xi)$.
Let $0\leq v(x)\leq1$. The PGFL of $\hat{\phi}_{r}(\xi)$ is given
by \begin{eqnarray}
\mathcal{G}_{\xi}(v(x)) & = & \mathbb{E}\prod_{x\in\phi_{r}}1-(1-v(x))\mathbf{1}(\xi\rightarrow x\mid\phi_{s})\label{eq:666}\\
 & \stackrel{(a)}{=} & \mathbb{E}\exp\left(-\lambda_{r}\int_{\mathbb{R}^{2}}(1-v(x))\mathbf{1}(\xi\rightarrow x\mid\phi_{s})dx\right)\nonumber \\
 & \stackrel{(b)}{\geq} & \exp\left(-\lambda_{r}\int_{\mathbb{R}^{2}}(1-v(x))p_{\xi}(x)dx\right),\label{eq:777}\end{eqnarray}
where $(a)$ follows since $\phi_{r}$ is a PPP, and $(b)$ follows
from Jensen's inequality. From the PGFL we observe that the point
process consisting of relays which connect to the origin is not a
PPP. $(b)$ would have been an equality if $\mathbf{1}(\xi\rightarrow x\mid\phi_{s})$
are independent for different $x$ and the resulting process would
be a PPP. But for the sake of analysis, we make the following assumptions
and justify them by simulations.
\begin{enumerate}
\item We assume that the spatial distribution of $\hat{\phi}_{r}(\xi)$
is an inhomogeneous PPP with intensity $\lambda_{r}p_{\xi}(\xi)$.
Since $T>1$, $\hat{\phi}_{r}(\xi_{1})\cap\hat{\phi}_{r}(\xi_{2})=\emptyset,\quad\forall\xi_{1},\xi_{2}\in\phi_{s}$. 
\item We also assume $\hat{\phi}_{r}(\xi_{1})$ is independent of $\hat{\phi}_{r}(\xi_{2})$
for all $\xi_{1},\xi_{2}\in\phi_{s}$.
\end{enumerate}
We will show the results obtained by this assumption are close to
the actual by simulation. From Figure \ref{Flo:theory-comp} we observe
that the simulation results (for Method $3$) are very close to that
predicted by theory making the above assumptions. This is intuitive
since many terms in \eqref{eq:666} are independent and thus the bound
in \eqref{eq:777} is very tight.

A subset of relays $N_{\xi}\subseteq\hat{\phi}_{r}(\xi)$ for each
$\xi\in\phi_{s}$ transmit in the second hop depending on the relay
selection method. This is basically a thinning of the point process
$\hat{\phi}_{r}(\xi)$. We will now derive the intensity of the point
process $N_{\xi}$ for different methods. We will denote the spatial
intensity of $N_{\xi}$ by $\Delta_{\xi}(z)$ and we have $\mathbb{E}[|N_{\xi}\cap A|]=\int_{A}\Delta_{\xi}(z)dz$
for any $A\subset\mathbb{R}^{2}$ .\\
Method $1$: Since $N_{\xi}=\hat{\phi}_{r}(\xi)$, we have $\Delta_{\xi}(z)=\lambda_{r}p_{\xi}(z)$.\\
Method $2$: From \eqref{eq:method21}, we have \begin{eqnarray*}
\Delta_{\xi}(z) & = & \lambda_{r}\mathbb{E}\left(\exp\left(-\delta\frac{\text{RSS}(z)}{1+T}\right)\mathbf{1}\left(\xi\rightarrow z\mid\phi_{s}\right)\right)\\
 & \stackrel{(a)}{=} & \lambda_{r}(1+T)\frac{\exp\left(-\lambda_{s}\int_{\mathbb{R}^{2}}\frac{(\delta g(\xi-z)+T)g(y)}{g(\xi-z)+(\delta g(\xi-z)+T)g(y)}dy\right)}{1+T+\delta g(\xi-z)},\end{eqnarray*}
where $(a)$ follows from a procedure similar to the evaluation of
$p_{\xi}(z)$. \\
Method $3$: Given $\xi$ and $r(\xi)$, we have \[
\Delta_{\xi}(z)=\lambda_{r}p_{\xi}(z)\mathbf{1}(\angle z\xi r(\xi)<\theta).\]
\\
Method $4$: Given $\xi$ and $r(\xi)$, we have\[
\Delta_{\xi}(z)=\lambda_{r}p_{\xi}(z)\exp\left(-\epsilon\|z-r(\xi)\|R^{-1}\right)\]
The average number of relays in a cluster $N_{\xi}$ that transmit
in the second hop is $\int\Delta_{\xi}(z)dz$.

\textit{Second hop: }The transmitting set in second hop is given by
$\psi=\bigcup_{\xi\in\phi_{s}}N_{\xi}$. Since $T>1$, at most one
transmitter belonging to $N_{\xi}$ can connect to $r(\xi)$. So the
probability that no node from $N_{\xi}$ can connect to $r(\xi)$
denoted by $f(\xi)$ is given by \begin{eqnarray*}
f(\xi) & = & 1-\sum_{z\in N_{\xi}}\mathbf{1}\left(z\rightarrow r(\xi)\mid\psi\right).\end{eqnarray*}
 So $\tilde{P_{2}}$ (same as $P_{2}$ without the limit) is given
by \begin{eqnarray*}
\tilde{P_{2}} & = & \mathbb{E}\sum_{x\in\phi_{t}\cap B(o,a)}\sum_{z\in N_{\xi}}\mathbf{1}\left(z\rightarrow r(\xi)\mid\psi\right)\\
 & \stackrel{(a)}{=} & \lambda_{t}\int_{B(o,a)}\mathbb{E}\sum_{z\in N_{\xi}}\mathbf{1}\left(z\rightarrow r(\xi)\mid N_{\xi}\cup\psi\right)d\xi\\
 & \stackrel{(b)}{=} & \lambda_{t}\int_{B(o,a)}\int\mathbb{E}\left[\Delta_{\xi}(z)\mathbf{1}\left(z\rightarrow r(\xi)\mid N_{\xi}\cup\psi\right)\right]dzd\xi\end{eqnarray*}
where $(a)$ and $(b)$ follow from the Campbell-Mecke theorem and
Slivnyak's theorem. We included $\Delta_{\xi}(x)$ in the expectation
operator because in Methods $3$ and $4$, $\Delta_{\xi}(z)$ depends
on the random variable $r(\xi)$. We now show that the inner integral
does not depend on $\xi$. We also have $\Delta_{\xi}(z)=\Delta_{o}(\xi-z)$
for Method $1$ and $2$. For Methods $3$ and $4$ we have $\Delta_{\xi}(z)\stackrel{d}{=}\Delta_{o}(\xi-z)$
where the equality is in distribution. Using the substitution $z'\leftarrow z-\xi$,
the stationarity of $\psi$, and the above property of $\Delta_{\xi}(z)$
we have \begin{equation}
P_{2}=\frac{1}{2\pi}\int_{0}^{2\pi}\int_{\mathbb{R}^{2}}\mathbb{E}\left[\Delta_{o}(z)\mathbf{1}\left(z\rightarrow R_{\nu}\mid\psi\cup N_{o}\right)\right]dzd\nu\label{eq:angle}\end{equation}
 where $R_{\nu}=(R\cos(\nu),R\sin(\nu))$. For Methods $1$ and $2$
by the isotropic nature of $N_{o}$ and $\psi$ we have \begin{equation}
P_{2}=\int_{\mathbb{R}^{2}}\Delta_{o}(z)\mathbb{E}\mathbf{1}\left(z\rightarrow R\mid\psi\cup N_{o}\right)dz\label{eq:method21}\end{equation}
Due to space constraints we will only describe how to derive $P_{2}$
for Method $3$. Method $4$ can be analyzed in a similar fashion.
From \eqref{eq:angle} and the definition of $\Delta_{o}(z)$ for
Method $3$, we have $P_{2}=$ \[
\frac{\lambda_{r}}{2\pi}\int_{0}^{2\pi}\int p_{o}(z)\mathbb{E}\mathbf{1}(\angle zoR_{\nu}<\theta)\mathbf{1}\left(z\rightarrow R_{\nu}\mid\psi\cup N_{o}\right)dzd\nu\]
 Since $\psi$ is isotropic we have \begin{equation}
P_{2}=\frac{\lambda_{r}\theta}{\pi}\int p_{o}(z)\mathbb{E}\mathbf{1}\left(z\rightarrow R\mid\psi\cup\tilde{N}_{o}\right)dz.\label{eq:MAIN-ERROR}\end{equation}
Here $\tilde{N}_{o}$ is equal to $N_{o}\cap S(o,R,\theta)$ where
$S(o,R,\theta)$ denotes a sector of angle $\theta$ on either side
of the line joining the origin and $(R,0)$. With a slight abuse of
notation we will denote $\tilde{N}_{o}$ also by $N_{o}$ and $N_{o}$
depends on the relay selection method. We now evaluate $\mathbb{E}\left[\mathbf{1}\left(z\rightarrow R\mid\psi'\right)\right]$
where $\psi'=\psi\cup N_{o}$. \begin{eqnarray}
\mathbb{E}\left[\mathbf{1}\left(z\rightarrow R\mid\psi'\right)\right] & = & \mathbb{P}\left(h_{zR}g(z-R)>TI(\psi',R)\right)\nonumber \\
 & \stackrel{(a)}{=} & \mathbb{E}\exp\left(-\frac{TI(\psi',R)}{g(z-R)}\right)\label{eq:3567}\end{eqnarray}
where $(a)$ follows from the exponential distribution of $h_{zR}$.
Also \eqref{eq:3567} is the Laplace transform of the interference
evaluated at $T/g(z-R)$. By our assumptions $\psi'$ is a Poisson
cluster process \cite{alsilomar2006,stoyan-1983} with an additional
cluster at the origin. The Laplace transform of the interference in
this case is given by $\mathbb{E}\exp\left(-s\sum_{y\in\psi'}h_{yR}g(y-R)\right)$
which is equal to\[
\mathbb{E}\left[\prod_{y\in\psi'}\exp\left(-sh_{yR}g(y-R)\right)\right]\stackrel{(a)}{=}G_{\psi'}\left(\frac{1}{1+sg(y-R)}\right)\]
where $G_{\psi}(.)$ is the PGFL of the process $\psi'$ and $(a)$
follows by Laplace transform of the fading. So we have \begin{eqnarray*}
 &  & \mathbb{E}\left[\mathbf{1}\left(z\rightarrow R\mid\psi'\right)\right]\\
 & \stackrel{(a)}{=} & \mathbb{E}\left[\prod_{\xi\in\phi_{s}}\prod_{y\in N_{\xi}}\frac{1}{1+\frac{Tg(y-R)}{g(z-R)}}\right]\mathbb{E}\left[\prod_{y\in N_{o}}\frac{1}{1+\frac{Tg(y-R)}{g(z-R)}}\right]\\
 & \stackrel{(b)}{=} & \mathbb{E}\left[\prod_{\xi\in\phi_{s}}\exp\left(-\tilde{\beta}(z-R,\xi)\right)\right]\exp\left(-\tilde{\beta}(z-R,R)\right)\\
 & \stackrel{(c)}{=} & \exp\left(-\tilde{\beta}(z-R,R)-\lambda_{s}\int1-\exp\left(-\tilde{\beta}(z-R,\xi)\right)d\xi\right),\end{eqnarray*}
where $(a)$ follows from assumption $2$, $(b)$ follows from a technique
similar to the derivation of $p_{\xi}(z)$, $(c)$ follows from the
PGFl of PPP, and where \[
\tilde{\beta}(z,\xi)=\int_{\mathbb{R}^{2}}\beta(z,y)\tilde{\Delta}_{o}(y+\xi)dy.\]
From the above equation, \eqref{eq:MAIN-ERROR} and \eqref{eq:method21}
we have, \begin{eqnarray*}
 &  & P_{2}=\int_{\mathbb{R}^{2}}\tilde{\Delta}_{o}(z+R)\\
 &  & \cdot\exp\left(-\tilde{\beta}(z,R)-\lambda_{s}\int_{\mathbb{R}^{2}}1-\exp\left(-\tilde{\beta}(z,\xi)\right)d\xi\right)dz,\end{eqnarray*}
 where $\tilde{\Delta}_{o}(z)=\Delta_{0}(z)$ for Method $1$ and
$2$. For Method $3$, $\tilde{\Delta}_{o}(z)=\frac{\theta}{\pi}\lambda_{r}p_{\xi}(z)$.
For Method $4$ \[
\tilde{\Delta}_{o}(z)=\frac{\lambda_{r}}{2\pi}p_{\xi}(z)\int_{0}^{2\pi}\exp\left(-\epsilon\|z-R_{\nu}\|R^{-1}\right)d\nu.\]

\bibliographystyle{plain}
\bibliography{point_process}

\end{document}